\documentclass [aps,twocolumn,showpacs,showkeys,superscriptaddress]{revtex4-1}

\usepackage{amssymb}
\usepackage{amsbsy}
\usepackage{amsmath}
\usepackage{amsfonts}
\usepackage{mathrsfs}
\usepackage{latexsym}
\usepackage[english]{babel}
\usepackage{url}
\usepackage{color}
\usepackage[]{graphicx}
\usepackage{epstopdf}
\usepackage{amsmath}
\usepackage{amssymb}
\usepackage{hyperref}

\begin{document}

\title{Experimental and theoretical evidences for the ice regime in planar artificial spin ices}

\author{R. P. Loreto}
\affiliation{Laboratory of Spintronics and Nanomagnetism ($LabSpiN$), Departamento de F\'{i}sica,
Universidade Federal de Vi\c{c}osa, 36570-000 - Vi\c{c}osa - Minas
Gerais - Brazil.}

\author{F.S. Nascimento} \affiliation{Departamento de F\'{i}sica,
Universidade Federal de Ouro Preto, 35931-008 - Jo\~{a}o Monlevade - Minas Gerais - Brazil.}

\author{R. S. Gonçalves}
\affiliation{Laboratory of Spintronics and Nanomagnetism ($LabSpiN$), Departamento de F\'{i}sica,
Universidade Federal de Vi\c{c}osa, 36570-000 - Vi\c{c}osa - Minas
Gerais - Brazil.}

\author{J. Borme}
\affiliation{INL-International Iberian Nanotechnology Laboratory, 4715-330, Braga, Portugal}

\author{J. C. Cezar}
\affiliation{Laborat\'{o}rio Nacional de Luz S\'{i}ncrotron-LNLS, CP 6192, 13083-970 Campinas, Brazil}

\author{C. Nisoli}
\affiliation{Theoretical Division and Institute for Materials Science, Los Alamos National Laboratory, Los Alamos, New Mexico 87545, USA}

\author{A. R. Pereira}
\email{apereira@ufv.br.}
\affiliation{Laboratory of Spintronics and Nanomagnetism ($LabSpiN$), Departamento de F\'{i}sica,
Universidade Federal de Vi\c{c}osa, 36570-000 - Vi\c{c}osa - Minas
Gerais - Brazil.}

\author{C.I.L. de Araujo}
\email{dearaujo@ufv.br}
\affiliation{Laboratory of Spintronics and Nanomagnetism ($LabSpiN$), Departamento de F\'{i}sica,
Universidade Federal de Vi\c{c}osa, 36570-000 - Vi\c{c}osa - Minas
Gerais - Brazil.}

\date{\today}

\begin{abstract}
In this work, we explore a kind of geometrical effect in the
thermodynamics of artificial spin ices ($ASI$). In general, such
artificial materials are athermal. Here, We demonstrate that
geometrically driven dynamics in $ASI$ can open up the panorama of
exploring distinct ground states and thermally magnetic monopole
excitations. It is shown that a particular $ASI$ lattice will
provide a richer thermodynamics with nanomagnet spins experiencing
less restriction to flip precisely in a kind of rhombic lattice.
This can be observed by analysis of only three types of rectangular
artificial spin ices ($RASI$). Denoting the horizontal and vertical
lattice spacings by $a$ and $b$, respectively, then, a $RASI$
material can be described by its aspect ratio $\gamma \equiv a/b$.
The rhombic lattice emerges when $\gamma = \sqrt{3}$. So, by
comparing the impact of thermal effects on the spin flips in these
three appropriate different $RASI$ arrays, it is possible to find a
system very close to the ice regime.

\end{abstract}

\flushbottom
\maketitle
%
%
\thispagestyle{empty}

\section*{Introduction}

Arrays of nanomagnets designed to resemble the spin ice materials
(disordered magnetic states) are known as artificial spin ices ($ASI$).
Nowadays, with the advances of the nanotechnology and nanofabrication,
$ASI$ systems have become so famous as well as their natural
counterparts, with the advantage that they can be constructed with
desirable geometries and properties. The first $ASI$ was built in $2006$
and it consists of a two-dimensional ($2d$) square array of $80,000$
elongated magnetic nanoislands, each a few hundred nanometers
long\cite{Wang2006}. The net magnetic moment (spin) of each individual
nanoisland is aligned parallel to its longest axis (like in a
bar magnet), and is coupled to all other nanoislands of the planar
array by the ubiquitous dipolar interaction. Then, in its original
configuration, $ASI$ tiles a square lattice of vertices, with four
nanoislands meeting at each vertex.

The ground state of the artificial square ice obeys the ice rule,
which remains the familiar two-in, two-out (two spins must point in,
while the other two must point out in each vertex). However, in two
dimensions, the standard ice rule is no longer
degenerate\cite{Wang2006,Moller2006,Mol2009}. In addition, $ASI$
systems have long been athermal (these compounds were almost always
found in frozen at room temperature), until the most recent
investigations on patterned ultrathin magnetic films could pave the
way to explore and visualize the real-time dynamics of all kinds of
different frustrated geometries. Indeed, recently, several works
have given attention to certain thermal properties of $ASI$
compounds in diverse types of planar
lattices\cite{Morgan2011,Silva12,Kapaklis,Farhan13,Kapaklis14,Farhan14,Farhan3,Thonig14,Andersson16}.
However, the $2d$ lattice obeying the usual two-in, two-out ice rule
with a degenerated ground state did not deserve yet an adequate
treatment. In this paper, we would like to do this by comparing
arrays that exhibit different ground states, when they are heated to
a high temperature regime (we mean by high temperature, the
practical values in the range $300 K-800 K$). Some of their magnetic
properties (as a function of temperature $T$) are observed by
Photoemission Electron Microscopy ($PEEM$) combined with $X$-ray
Magnetic Circular Dichroism ($XMCD$) measurements. $MOKE$ signals
are also analyzed. It is done by deforming continuously the square
lattice in a rectangular one. Therefore, we experimentally focus on
rectangular artificial spin ices ($RASI$) with horizontal and
vertical lattice parameters given by $a$ and $b$ respectively ($b$
is kept constant while $a$ is varied). Our samples match in three
classes of $RASI$, appropriately designed to illustrate how a
particular planar system can approximate of the ice regime. They are
heated from room temperature until a temperature near $800 K$ (just
below the Curie Temperature of the permalloy, which is $T_{C}=873
K$).
\begin{figure}
    \centering
    \includegraphics[width=8.5 cm]{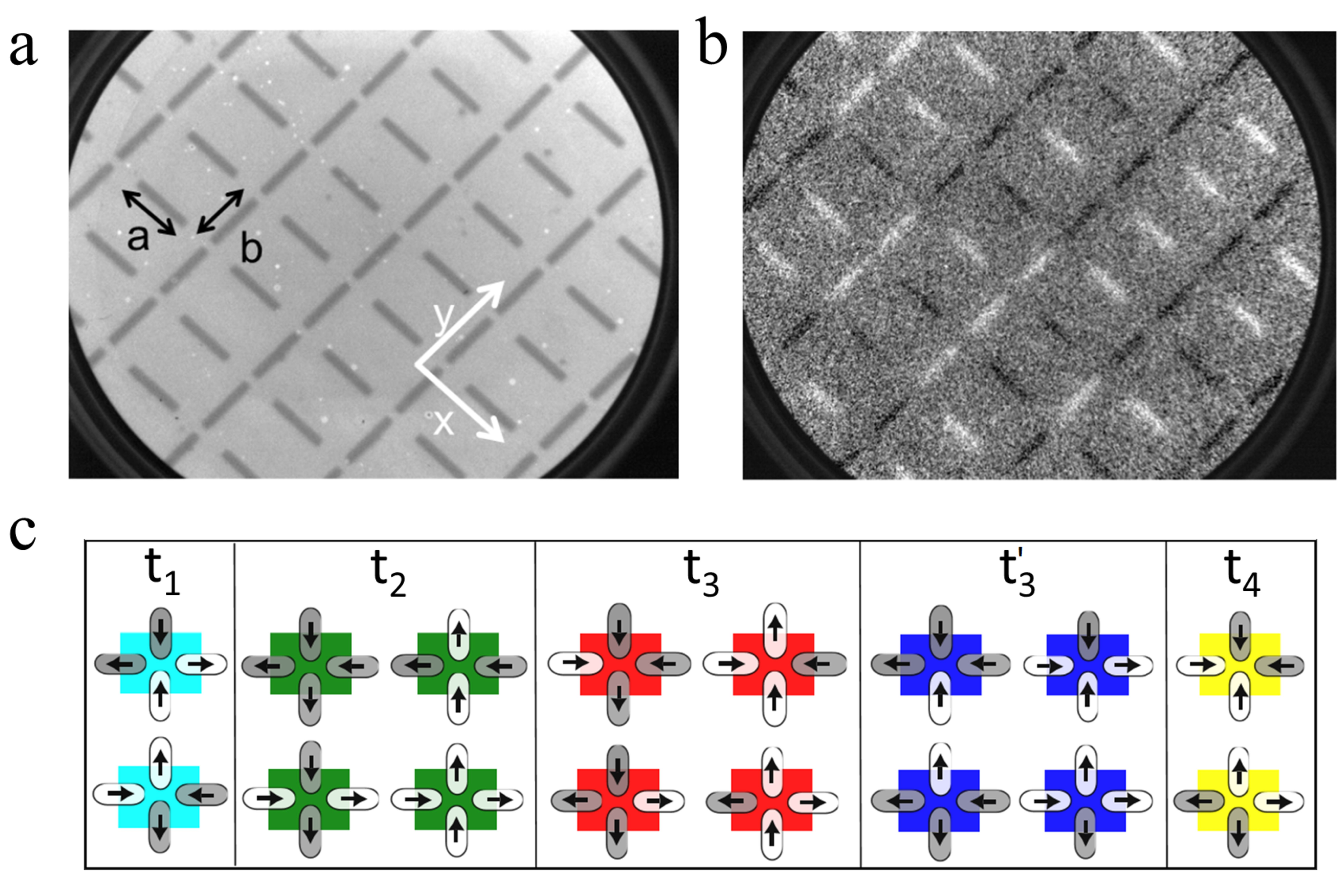}
    \caption{ \textbf{(a)} $30 \mu$m x $30 \mu$m $PEEM$ image of $\gamma = \sqrt{3}$ $RASI$ lattice. The Permalloy nanoislands
    have $2800 nm x 400 nm x 10 nm$. It would be useful to say that those nanomagnets, larger then usual $ASI$ systems, were used
    in order to achieve higher image contrast in our $PEEM$, after careful tests performed by micromagnetic simulations and
    Magnetic Force Measurements, which ensured that it is inside the limit to present monodomains (not supporting vortices or
    multidomains). \textbf{(b)} Typical $XMCD$ measurement for the same area of (a) with the clear and dark contrast representing
    the orientation of the islands monodomains, in each direction. The measurement was taken with the array rotated $45$ degrees
    from the $X$-ray sensitivity to resolve $x$ and $y$ direction at the same time. \textbf{c)} Possible vertex types for each vertex represented
    by different colors, where arrows represent the orientation of the island related to $XMCD$ pattern.}
    \label{fig:fig1}
\end{figure}

Theoretical calculations\cite{Nascimento12} concerning the
rectangular artificial systems suggest that the ice regime (with the
required degenerate ground state), could be observed when the aspect
ratio of the lattice ($\gamma \equiv a/b$) is equal to $\sqrt{3}$
(the rhombic lattice). On the other hand, like the artificial square
array, the ground states of $RASI$ compounds with $1 < \gamma <
\sqrt{3}$ and $\gamma > \sqrt{3}$ are not degenerate, but they have
very distinct magnetic properties: in the first case ($1 < \gamma <
\sqrt{3}$), there are residual magnetic charges at each vertex,
alternating from positive to negative along the neighbor vertices
(these charges are not cylindrically symmetric exhibiting a strong
quadrupole moment). This system can be characterized as
antiferromagnetic (along the vertical and horizontal lines of
spins). In the second case ($\gamma > \sqrt{3}$), there are
alternating residual magnetic moments along the neighbor vertices.
Again, looking the vertical and horizontal lines of spins, it can be
characterized as a ferromagnetic state. At $\gamma = \gamma_{R} =
\sqrt{3}$, these two distinct types of states have the same energy
and the ground state becomes degenerated. This special case
separates the antiferromagnetic state ($1 < \gamma < \sqrt{3}$) from
the ferromagnetic one ($\gamma > \sqrt{3}$) and can be distinguished
as a more realistic spin ice state in these artificial systems
(later, the difficulty of obtaining an accurate ice state,
theoretically and experimentally, will be discussed).

Figure 1a shows an image of a $RASI$ with $\gamma =\sqrt{3}$
obtained by $PEEM$ measurements combined with $XMCD$ technique (see
also Fig. 1b), displaying the magnetic monodomains of the
nanoislands. Figure 1c shows the five possible vertex types of these
structures. Vertex types $t_{1}$ and $t_{2}$ obey the ice rule
(two-in, two-out) while the other three ($t_{3}, t'_{3}$ and
$t_{4}$) represent excited states singly ($t_{3}$ and $t'_{3}$) and
doubly ($t_{4}$) charged magnetic monopoles\cite{Nascimento12}. The
energy of all these vertex types depends on $\gamma$. For $\gamma =
\gamma_{R}$, the $t_{1}$ and $t_{2}$ types have the same energy,
yielding to a degenerate ground state. To easier see such an array,
it is better to replace the net magnetic moment of the nanoislands
by a point-like dipole at their centers. Then, the four dipoles
(spins) for the case $\gamma=\sqrt{3}$ are located at the vertices
of a rhombus with short and long diagonals $b$ and $a=\sqrt{3}b$,
respectively and, therefore, they are equidistant (in this case, the
distance of each pair of spins in a rhombus is $b$). Then, we can
say that the spins are placed at the vertices of a kind of rhombic
lattice with rhombi having spins pointing out parallel to their
diagonals alternating (along the diagonal of the rectangles) with
rhombi having spins pointing out perpendicular to their diagonals. A
rhombic lattice is one of the five $2d$ lattice types as given by
the crystallographic restriction theorem.

Here, we show theoretically and experimentally that the geometry
distortion of the planar arrays may be an additional ingredient able
to cause important physical phenomena in $ASI$ materials. In fact,
by studying three $RASI$ systems with different aspect ratios
$\gamma$, we demonstrate that the geometrical influence goes beyond
the simple effect of the variation of the lattice parameters. To
explain the aims of this work, we organize the sequence of the paper
as follows: firstly, we use Monte Carlo ($MC$) simulations to
calculate the specific heat and topologies density as functions of
temperature $T$. In principle, based on the square lattice, these
thermodynamic quantities have important features only when the
temperature scale is on the order of $10^{4} K$. If it is true, most
of these theoretical results could not be experimentally verified
but they will give us, at least, interesting insights about $ASI$
and $RASI$. On the other hand, our $MC$ simulations also suggest
that the critical temperatures for the specific heat fall rapidly as
$\gamma$ increases till $\sqrt{3}$, guiding some possibilities and
questions about $RASI$ that only experiments could answer. Hence, an
experimental investigation is the second natural step of this paper.
We then experimentally measure the magnetization and topology
density as functions of $T$ by heating the samples from room
temperature ($300 K$) to $\sim 750 K$. Although this range of
temperature is about ($10-100$) times smaller than the typical
temperature scale necessary to easily flip the large spin of the
magnetic nanoislands, we can directly observe very interesting
thermal effects in some artificial materials, when the geometry is
stretched from the square to the rectangular lattice. The main
conclusion here is that arrays with $\gamma \approx 1$ (almost
square) are essentially athermal systems as already mentioned in
several works, while arrays with $\gamma \approx \sqrt{3}$ present
important thermal and dynamical effects even at a temperature scale
on the order of $10^{2}-10^{3} K$. We will see that this effect does
not have to do with the fact that the rectangular arrays (with
$\gamma > 1$) are more weakly coupled than the square array (of side
$b$). Really, although the temperature-induced onset of magnetic
fluctuations increases with the lattice spacing and related
interaction strength between nanoislands\cite{Kapaklis14}, our
results show that an extra factor will be more fundamental here:
lattice geometry. A purely geometrical effect will make the net
magnetic moments to appear flexible enough to rather flip with a
smaller amount of effort (for temperatures compatible with
artificial magnets), allowing the dipolar interaction to have more
protagonism in determining the dynamical properties of the system
than usually it does. For investigating the two aims cited above,
$RASI$ materials with aspect ratios equal to $\gamma =\sqrt{2}$,
$\gamma= \gamma_{R} = \sqrt{3}$ and $\gamma = \sqrt{4}$ were
appropriately chosen and fabricated. So the effects proposed here
will be exposed by comparing some thermodynamic quantities of these
three rectangular arrays, observing the roles of their geometries.

\begin{figure}
    \centering
    \includegraphics[width=8.5 cm]{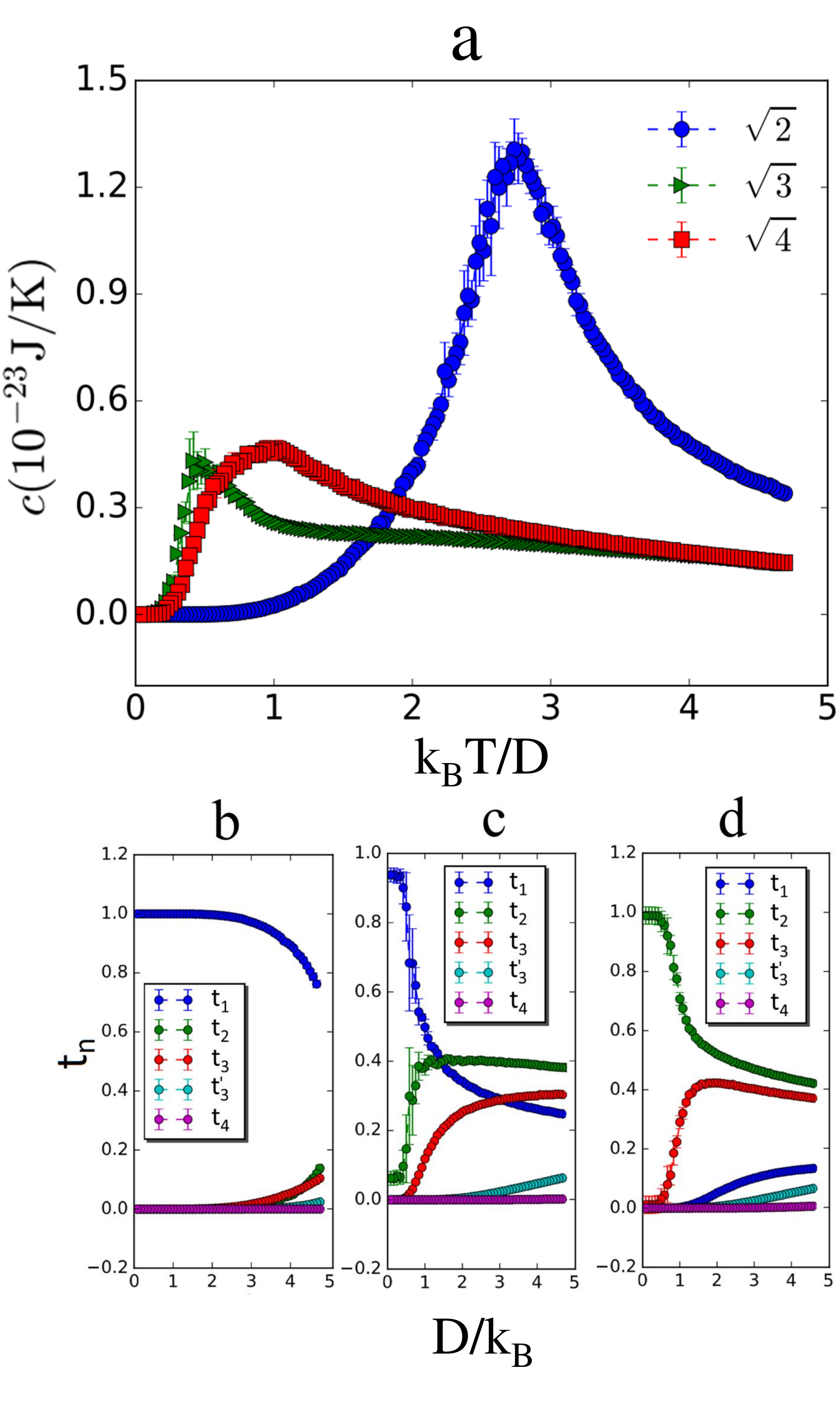}
    \caption {\textbf{(a)} Specific heat of $RASI$ with different
    values of $\gamma$ obtained by Monte Carlo simulations.
    Parts \textbf{(b)}, \textbf{(c)} and \textbf{(d)} show the density
    of states for each vertex type ($t_{1},t_{2},t_{3}, t'_{3}$ and
    $t_{4}$) in the lattices with $\gamma = \sqrt{2}$, $\gamma = \sqrt{3}$
    and $\gamma = \sqrt{4}$ respectively, as a function of temperature. }
    \label{fig:fig2}
\end{figure}

\section*{Results}

We start by reporting the thermodynamic results for $RASI$ as
obtained by $MC$ simulations. The calculations presented here use
the point-dipole approximation but we have also done calculations
using the dumbbell model. We did not find any relevant differences
between the two methods. The model is implemented by standard $MC$
techniques on a system with $N=364$ spins. Here, the spin-spin
interaction is assumed to be purely dipole-dipole such that the
Hamiltonian is given by $H=D\sum_{i>j} \left[
\frac{\hat{e}_i\cdot\hat{e}_j}{r^2_{ij}}-\frac{3(\hat{e}_i\cdot\vec{r}_{ij})(\hat{e}_j\cdot\vec{r}_{ij})}{r^5_{ij}}
\right] s_i s_j$, where $D\approx2.1\times10^{-19}\mathrm{J}$ is the
coupling constant of the dipolar interaction, $\hat{e}_i$ is the
local Ising axes of the rectangular lattice, $r_{ij}$ is the
distance between spins and $s_i=\pm1$ represents the two states
(up/down) of the Ising spin. In our procedure, Monte Carlo step
consists of $N$ single-spin flips and we have used $10^4$ Monte
Carlo steps to reach equilibrium configurations and $10^5$ steps to
get thermodynamics averages. Samples are first prepared in a
disordered state and then cooled to low temperatures; this annealing
protocol in general drives the spin configuration to the ground
state. In this process, the specific heat is calculated by the
fluctuations in the total energy of the system, $c=(\Delta E)^2/k_B T^2 N$.

For the usual artificial square ice, previous calculations by Silva
\textit{et al.}\cite{Silva12} have suggested the existence of a
phase transition at a critical temperature $ T_{P} \sim 7.2
D/k_{B}$, where $D$ is the coupling constant of the dipolar
interaction and $k_{B}$ is the Boltzmann constant. Really, the
specific heat exhibits a sharp peak at $T_{P}$, whereas the
amplitude diverges logarithmically with the system size $L$. Such a
phase transition was speculated to be attributed to the vanishing of
the string tension joining monopoles of oppositive charges: below
$T_{P}$, the monopoles are joined by an energetic (and observable)
string (Nambu monopoles\cite{Nambu74}); above $T_{P}$, the string
tension should vanish and some monopoles become free to move
(actually, they may not be completely free because a monopole pair
is subjected to an entropic force that exhibits, in two dimensions,
a logarithmic distance dependence\cite{Silva12,Moler09}). Concerning
the specific heat, our $MC$ calculations lead to similar behaviors
for the $RASI$ materials. Figure 2 shows the specific heat as a
function of temperature for the three different cases considered
here. As expected, the critical temperature ($T_{\sqrt{n}}$,
$n=2,3,4$), at which the peaks occur, is a function of the aspect
ratio $\gamma$. Initially, as $\gamma$ is increased from $1$ (square
ice), the critical temperature decreases. It is expected because the
coupling among nanoislands becomes weaker as the array becomes more
stretched. However, interestingly, this critical temperature has a
minimum for $\gamma=\sqrt{3}$, in such a way that $T_{P} >
T_{\sqrt{2}} > T_{\sqrt{3}} < T_{\sqrt{4}} < T_{\sqrt{2}}$. Except
for $\gamma=\sqrt{3}$, the reasonable idea that the critical
temperature decreases as $\gamma$ increases works very well
($T_{\sqrt{2}}=2.7 D/k_{B}$, $T_{\sqrt{3}}=0.5 D/k_{B}$ and
$T_{\sqrt{4}}=0.8 D/k_{B}$) see Fig. 2a). It reinforces the fact
that a rhombic ice has special properties as demonstrated by
previous calculations\cite{Nascimento12}. Indeed, to our knowledge,
it is the only one planar case, obeying the familiar two-in, two-out
ice rule, that has a degenerate ground state (topologies $t_{1}$ and
$t_{2}$ possess the same energy). In these circumstances, the string
tension should be zero for any temperature (including the absolute
zero). However, an entropic effect generates an attractive
interaction potential ($T \ln R$, where $R$ is the distance between
a monopole and its antimonopole in a pair) in such a way that
monopoles should become free only at $T=0$. For finite temperatures,
they may be found apart but not completely free. To try to observe
some vestiges of the geometry features, we have also calculated the
topology densities as functions of temperature. These results are
shown in Figs.2b, 2c and 2d. The rectangular ice magnets with
$\gamma=\sqrt{2}$ and $\gamma = \sqrt{4}$ start (at $T=0$) with all
vertices in their respective natural ground states (the vertex types
$t_{1}$ and $t_{2}$, respectively). On the other hand, the rhombic
ice ($\gamma_{R}=\sqrt{3}$) starts with all vertices in the vertex
type $t_{1}$, which may be merely one of its possible ground states.
In all situations, as $T$ is increased from zero, the other vertex
types start slowly to arise around the lattice. It occurs in
different manners as $\gamma$ changes. For $\gamma = \sqrt{2}$, the
system practically remains entirely in its ground state for a large
range of temperatures, in such a way that, the the vertex type
$t_{1}$ begins to really decrease only for $T > 3 D/k_{B}$ (Fig.2b).
Like the square ice, this high temperature value for the
materialization of excitations indicates that systems with
relatively small aspect ratio ($\gamma \approx 1$) tend to be
athermal (in the context frequently used in the literature for
$ASI$). Conversely, for the special case of a rhombic lattice, the
initial state (with all vertices in $t_{1}$) reduces rapidly as $T$
increases and, simultaneously, the density of the $t_{2}$ vertex
type increases considerably (similar results are obtained if the
initial state would have all vertices in $t_{2}$, which is also a
possible ground state in this case). It suggests an immediate
activity in the spin fluctuations, even at relatively low
temperatures. Note that the density of the ground state vertex types
($t_{1}$ and $t_{2}$) becomes exactly equal just at the critical
temperature $T_{\sqrt{3}}$ (see Fig.2c). Only these two vertex types
are essentially fluctuating for low $T$. It means that, for $0 < T <
T_{\sqrt{3}}$, the system still persists almost entirely in its
ground states ($t_{1}$ and $t_{2}$ have the same energy when $\gamma
= \sqrt{3}$), since the vertex types $t_{3}, t'_{3}$ and $t_{4}$
were not significantly excited yet. Indeed, for this particular
case, any mixture of the states $t_{1}$ and $t_{2}$, if it is
possible to occur, would also be a ground state. Therefore, as the
density of $t_{2}$ increases (and the density of the excited states
$t_{3}, t'_{3}$ and $t_{4}$ remains close to zero), this degenerate
system seems to be accessing several of its different ground states.
For $T > T_{\sqrt{3}}$, the vertex types $t_{3}$ and $t'_{3}$ start
to be excited and then, this upper temperature phase rather
signalizes the emergence of monopoles with tensionless strings.
Finally, Fig.2d shows that the case $\gamma = \sqrt{4}$ has a
behavior qualitatively similar to the first example studied here
($\gamma = \sqrt{2}$). The basic differences are the ground states
and the fact that the $t_{3}$ monopoles (red circles) are more
easily excited in the array with $\gamma = \sqrt{4}$, producing a
lower transition temperature. The expressive presence of the $t_{3}$
monopoles in comparison with the $t'_{3}$ monopoles (in all cases)
is justified because they have smaller energy (actually, the
difference between the energy of $t'_{3}$ and $t_{3}$ vertex types
increases as $\gamma$ increases\cite{Nascimento12}). The above
theoretical calculations are very suggestive: $T_{\sqrt{3}}$ is
about $15$ times smaller than $T_{P}$, which means that the rhombic
ice exhibits, at much lower temperatures, a richer dynamics than the
square ice. Moreover, it is not only due to a larger mean distance
among the magnetic nanoislands, but mainly, due to the lattice
geometry itself since the $RASI$ with $\gamma = \sqrt{4}$ presents
less fluctuations than the artificial rhombic lattice. Therefore,
$RASI$ materials deserve a deeper investigation from the
experimental point of view. This is a natural next step: the study
of high temperature $RASI$. Here, high temperature means the
practical range in the interval $300 K < T < 800 K$.

\begin{figure*}
    \centering
    \includegraphics[width=15 cm]{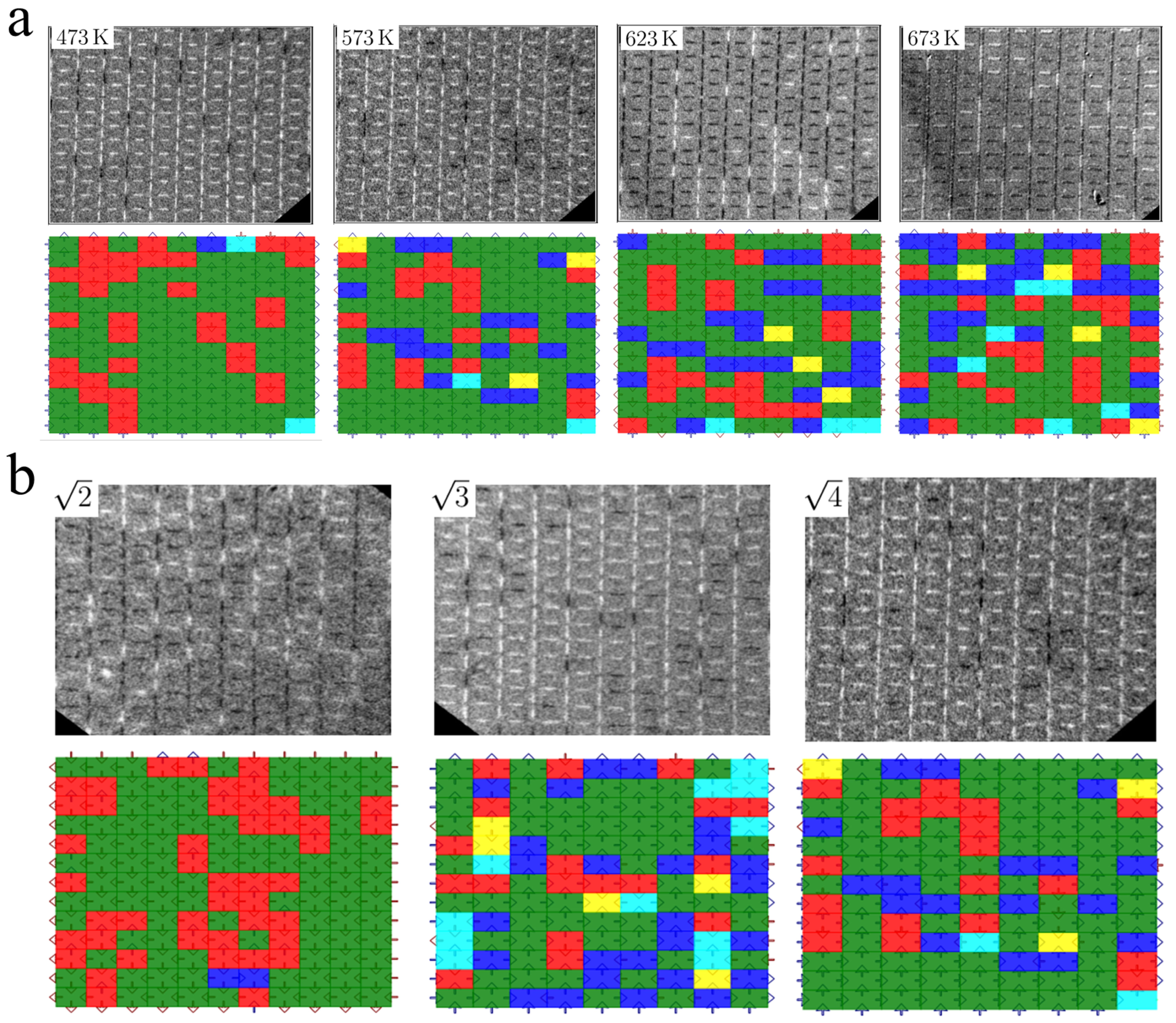}
    \caption {\textbf{(a)} 100 $\mu$m x 100 $\mu$m $PEEM-XMCD$ images of $\gamma = \sqrt{4}$ $RASI$ for different temperatures
    (up) and mapped respective magnetic configurations (down). \textbf{(b} $PEEM-XMCD$ images for
    $\gamma = \sqrt{2}$, $\gamma = \sqrt{3}$ and $\gamma = \sqrt{4}$ for same temperature of $573 K$}
    \label{fig:fig3}
\end{figure*}

With these indicative theoretical results in mind, we now consider realistic
experimental arrangements in which the temperature of the three
kinds of $RASI$ systems varies from $300 K$ to $750 K$. This range of $T$ is,
in principle, about $10-100$ times smaller than the temperature scale of the
most theoretical calculations presented above, but it is below the Curie
temperature ($T_{C}$) of the permalloy nanoislands.

For the fabrication of Permalloy nanoislands, a multilayer with
composition $Si$ / $Ta$ $3 nm$ / $Ni_{80} Fe_{20}$ $10 nm$ / $Ta$
$3nm$ was previously prepared by sputtering from tantalum (seed and
cap layer) and alloyed permalloy target, on silicon substrate. Then,
the samples were covered with a $85 nm$ layer of $AR-N 7520.18$
negative tone photoresist and patterned by electron lithography at
$100 kV$ of acceleration voltage. After development, the samples
were etched by ion milling at 20$^\circ$ from normal incidence,
using secondary ion mass spectroscopy to detect the end of the
process. An ashing in oxygen plasma was subsequently performed to
remove the photoresist. The nanoislands dimensions of $l = 2800 nm$
and $w = 400 nm$ were conceived in order to present magnetic
monodomain in each island. The $y$-axis lattice spacing of $b=3550
nm$ was kept in all samples and the $x$-axis lattice constant $a$
ranged from $5017-7100 nm$ in such a way that we have investigated
by ($PEEM-XMCD$), $RASI$ arrangements with aspect ratios
$a/b=\sqrt{2}, \sqrt{3}$ and $\sqrt{4}$ (see Fig. 3). For the
$PEEM-XMCD$ measurements the samples were heated at different
temperatures and images were taken just after temperature switch-off
to prevent different sample holder dilatation and beam defocusing.
The $MOKE$ signal was obtained during sample heating. These systems
were built on an area of $100\mu m^{2}$, which enabled topologies
density analysis in arrays of up $10 \times 14$ unit cell ($280$
islands). We have also done some Monte Carlo numerical calculations
to compare with experimental results.

The $XMCD$ measurements were performed at $PGM$ beamline of the
Brazilian Synchrotron Light Laboratory \footnote{lnls.cnpem.br/en/}.
They result from the core-level absorption of circularly polarized
soft $X$-ray by a magnetic element and the transfer of right ($RCP$)
or left ($LCP$) circularly polarized angular momentum of the photons
to the excited photoelectrons. The spin and orbital moments can be
determined from linear combinations of the dichroic difference
intensities of $RCP$ and $LCP$. The images were taken on the Nickel
$L_{2,3}$ edge with a photon energy of $850 eV$. The islands array
was placed rotated $45^{\circ}$ related to the $X$-ray sensitivity
in order to fully resolve both $\textit{x}$ and $\textit{y}$
directions. $MOKE$ measurements were made using a $p$-polarized
source. The islands array were placed rotated $45^{\circ}$ related
to the scattering plane to fully resolve both  $\textit{x}$ and
$\textit{y}$ directions. The array was saturated by an external
magnetic field and the samples were heated in a low vacuum
environment to prevent oxidation. The $MOKE$ signal was recorded by
varying the temperature in the sample.

Figure 3a shows a sequence of snapshots of a $RASI$ with $\gamma = \sqrt{4}$.
The images were obtained with samples initially at room temperature ($RT$) and
heated to $473 K$, $573 K$, $623 K$ and $673 K$. The ground state of this
array\cite{Nascimento12} should have all vertices in the vertex type $t_{1}$.
Figure 3a shows, more explicitly, how the temperature can actuate to generate
fluctuations, creating and destroying all possible low energy excitations into
the system (including monopole pairs). Note also that the density of these low
energy excitations increases with $T$. These pictures also demonstrate that most
of these excitations are displayed along the vertical direction, since this is
the case with less energy\cite{Nascimento12}. To emphasize and illustrate the
role of the thermal effects on the creation of excitations in $RASI$ materials,
we investigate experimentally how magnetized samples will be affected as the
temperature is increased. Figure 3b presents a comparison of the vertex types
for different $RASI$ ratio at the same temperature. Those measurements corroborate
the prediction of higher density of high energetic
vertex types for the orthorhombic $\gamma = \sqrt{3}$ geometry \cite{igor}.

\begin{figure}
    \centering
    \includegraphics[width=8.5 cm]{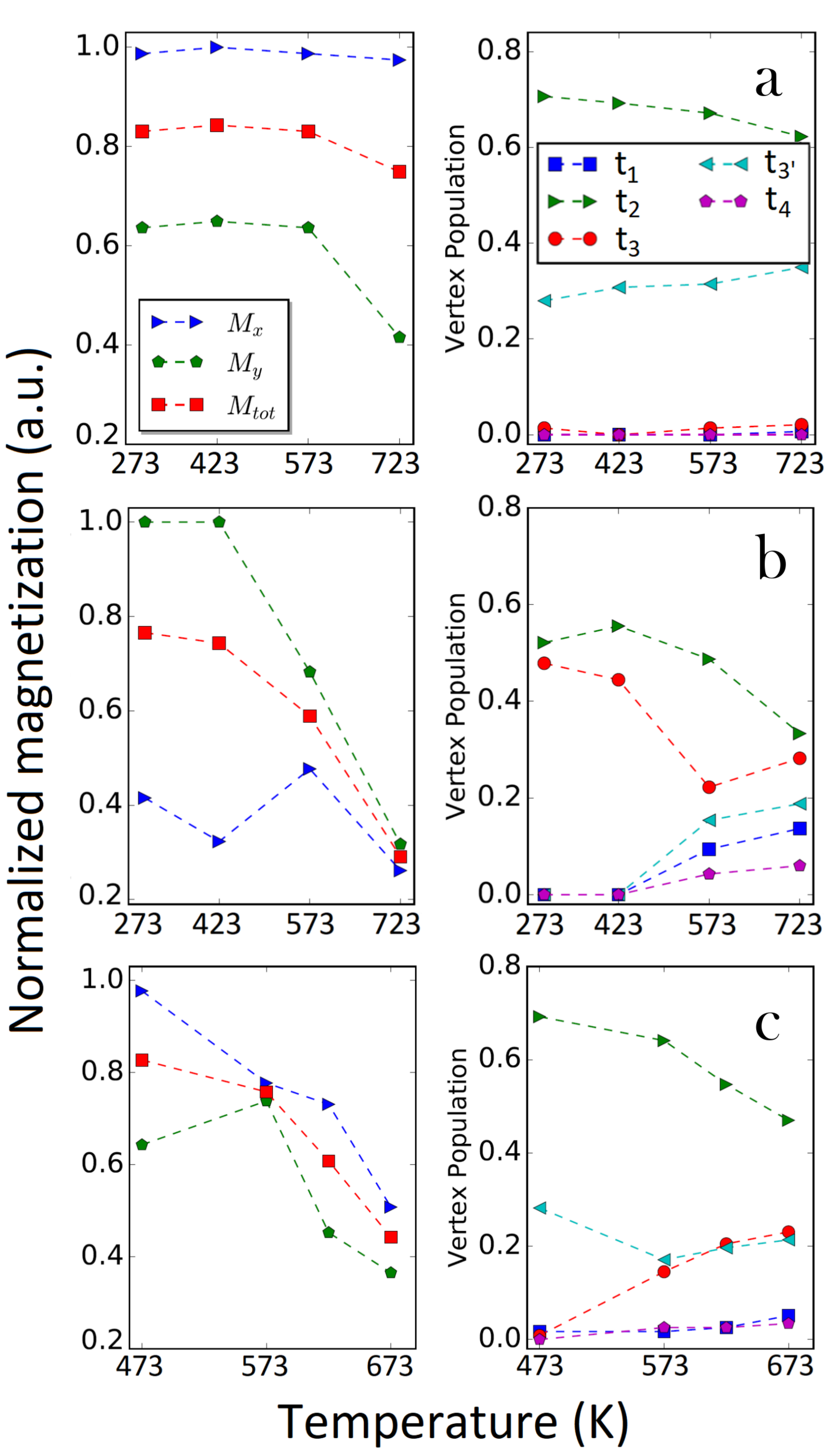}
    \caption{Normalized magnetization and population of vertex types for lattices with
    \textbf{(a)} $\gamma = \sqrt{2}$, \textbf{(b)} $\gamma = \sqrt{3}$ and \textbf{(c)}
    $\gamma = \sqrt{4}$. The magnetization and topologies are calculated from the $XMCD$
    measurements as shown in Fig. 3. }
    \label{fig:fig4}
\end{figure}

All samples were initially prepared (at $RT$) with normalized
magnetization along lattice diagonal. It was used four values for
$T$. As the temperature is increased, $M_{tot}$  decreases for the
three cases but with different behaviors. For instance, Fig.4a shows
the Magnetization (left) and the vertex types density (right) for
the case $\gamma = \sqrt{2}$ as the temperature is varied from $300
K$ to $\sim 750 K$. Note that the magnetization decreases very
slowly as $T$ increases. Meanwhile, there is a slow increasing of
the monopoles number ($t_{2}$ and $t_{3}$). Such a behavior reveals
that this system, like the square ice, is essentially athermal in
the range of temperatures compatible with experiments in artificial
spin ices. The magnetic moments of the nanoislands do not flip
easily and the magnetization remains practically unchanged during
all interval of temperatures studied. On the other hand, for the
special case of $\gamma = \sqrt{3}$, the magnetization decreases
very rapidly as $T$ increases (left side of Fig.4b) and seems to
vanish even at $T < T_{C}$, while the monopoles population
(including also doubly charged monopoles) increases considerably as
$T$ increases, mainly for $T > 550 K$ (right side of Fig.4b). This
example indicates that the geometry may produce favorable conditions
to flip the magnetic moments, making the system to exhibit its
thermodynamic properties at lower temperatures. To confirm this
fact, we have also investigated the case $\gamma =\sqrt{4}$. It
suggests that, for $\gamma > \gamma_{R}$, the magnetization
decreases somewhat slower again as $T$ increases (see the left side
of Fig.4c), but it falls much more rapidly as compared to the array
with $ \gamma = \sqrt{2}$. The different temperature range was chosen to better show the magnetization decay region, which is slightly different from the two other previous samples, once in this particular sample the
ground state is predicted to be in the ferromagnetic regime. We then notice that the experimental data obtained for the magnetization can be connected with the theoretical calculations for the specific heat: the temperature in which
$M_{tot}$ goes to zero and the specific heat presents a peak is a
minimum for $\gamma=\sqrt{3}$. These thermodynamical features are
caused mainly by geometrical effects, reinforcing the idea that the
string tension tends to vanish for the rhombic ice. This lattice is
singular since it changes the tendencies in rectangular arrays as
$\gamma$ is increased from $1$ (square ice). Indeed, if the lattice
were stretched continuously, one expects that the critical
temperature should decrease monotonically as $\gamma$ increases
because the mean distance among the nanoislands increases (and as a
consequence, the mean coupling among the spins decreases). Of
course, it is not the case here. Figure 5 shows the $in-situ$ $MOKE$
signal measured as a function of temperature in the lattice diagonal
direction, just after magnetization saturation in the same
direction. These results also confirm the geometrical deformation
effect observed in the $MC$ simulations and $XMCD$ measurements. The
difference in the slope of $\gamma=\sqrt{3}$ curve comes from the
fact that the ground state is degenerated and then the saturation,
which imply in $t_{2}$ vertex type, remains at low temperatures in
comparison with the other geometries that favors vertex
configuration to ground state or formation of high energy vertex
configuration with no residual magnetization.
\begin{figure}
    \centering
    \includegraphics[width=9cm]{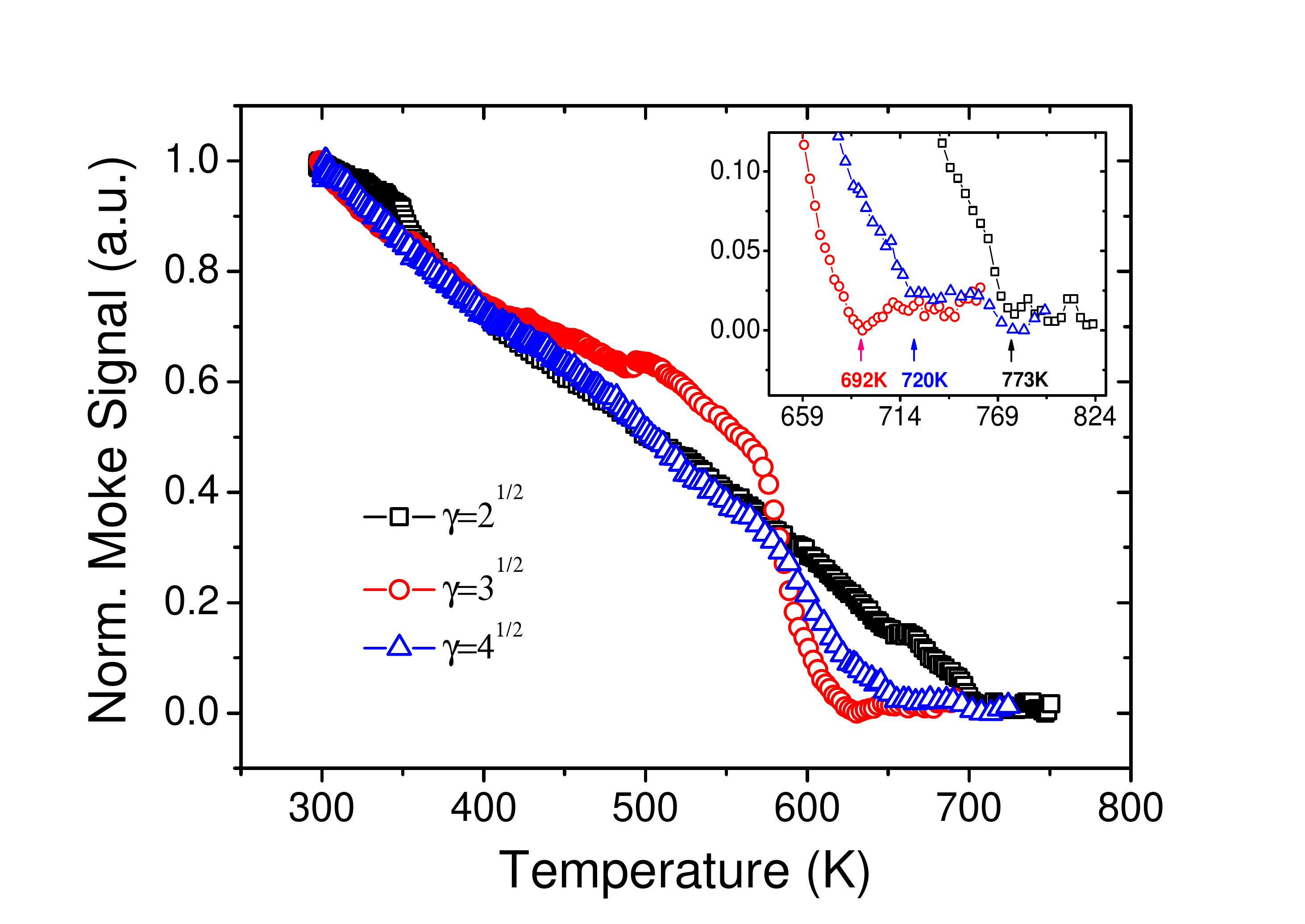}
    \caption{$MOKE$ signal as function of temperature for the three lattices compared.
    Here $T_{\sqrt{2}} \approx 707K$, $T_{\sqrt{3}} \approx 630K$ and $T_{\sqrt{4}} \approx 657K$.}
    \label{fig:fig5}
\end{figure}

\section*{Discussion and conclusion}

Important questions about the systems investigated here remain to be
explained. They are mainly related to the doubts about if it is
really possible to achieve the  perfect ice regime in these
artificial materials. We must say that if the system with aspect
ratio $\gamma_{R}=\sqrt{3}$ were really in the ice state, then there
would be no peak in the specific heat (see Fig.2) but just a smooth
bump, because there is no phase transition to the ice state, just a
crossover. If it is so, one should not say that this configuration
is really degenerate, but just that it is effectively degenerate for
$T > 0$. However, some details of the theoretical approach and
experimental measurements must be observed before any conclusion
about a desirable complete and precise description of these
artificial compounds. Indeed, $\gamma_{R}$ is not a rational number
and, therefore, any calculation of the specific heat is not done
exactly at the correct value in which the system degenerates. For
the simulations, the use of any rational number $r$ very close to
$\sqrt{3}$ will imply that the system is only near to the ice
regime, but it still keeps some features of the antiferromagnetic
(if $r$ is immediately below $\sqrt{3}$) or ferromagnetic phase (if
$r$ is immediately above $\sqrt{3}$). Experimentally, things are
still more complicated because the measurements of distances between
islands have the errors of the instruments and, in addition, the
nanoislands are not point dipoles. So, all results concerning the
lattices analyzed are only approximations. Despite these
difficulties, our results indicate (theoretically as well as
experimentally) that geometry may induce some roles on the thermodynamic
properties of these systems.

In conclusion, we have demonstrated experimentally that the lattice
geometry can be an important ingredient to transform the
thermodynamic properties of artificial spin ice compounds. By
stretching the lattice, the fluctuations do not decrease
monotonically as expected when the mean distances among the
nanoislands increase. Indeed, normally, $ASI$ materials are athermal
and the spins do not flip easily because a nanoisland contains a
large number of atoms. On the other hand, it has been shown that an
unambiguous stretching of the square ice may take the system to the
ice regime; there is a peculiar rectangular array with aspect ratio
$\gamma = \gamma_{R}=\sqrt{3}$ (rhombic ice), in which the
fluctuations become more evident as compared to its counterparts.
Note that the nanomagnets constituents of the artificial rhombic ice
are akin to the nanomagnets composing all others types of arrays
built for this investigation. Therefore, geometry must be important
for the different behaviors observed in these systems. The rhombic
lattice makes equidistant the four spins meeting at a vertex,
similar to the three-dimensional natural and even artificial spin
ices\cite{Rougemaille}. This geometric effect has shown to be more
effective in inducing spin fluctuations than the usual way of purely
increasing the lattice spacing among the nanoislands. Although all
the evidences obtained here suggest that the rhombic ice is in the
ice regime, we must be very careful in stating that. To be sure
about this, it should be useful to know the behavior of the magnetic
structure factor of the moment configurations. If it exhibits the
usual pinch singularities, we could assert that the spin ice
degeneracy has been really established. Nevertheless, such
calculations and/or neutron scattering experiments are out of the
scope of the present paper.

\section*{Acknowledgements}

The authors would like to thank the Brazilian agencies CNPq, FAPEMIG
and CAPES.

\end{document}